\documentclass{emulateapj}

\newcommand{\gray}{$\gamma$-ray}
\newcommand{\grays}{$\gamma$-rays}

\newcommand{\moon}{{\rm Moon}}
\newcommand{\pubbook}[6]{#1, #6, #2, #3, #4, #5}
\newcommand{\pubjournal}[6]{#1 #5, #2, #3, #4}
\newcommand{\pubjournala}[6]{#1 #5, #2, #4}

\newcommand{\geant}{GEANT4}

\newcommand{\icrc}{Int.\ Cosmic Ray Conf.\ }

\newcommand{\app}{APh}

\shorttitle{}
\shortauthors{Moskalenko et al.}

\begin{document}

\title{
A celestial gamma-ray foreground due to the albedo of small solar system bodies and
a remote probe of the interstellar cosmic ray spectrum
}

\author{Igor V. Moskalenko\altaffilmark{1}}
\affil{
   Hansen Experimental Physics Laboratory, 
   Stanford University, Stanford, CA 94305
\email{imos@stanford.edu}}
\altaffiltext{1}{Also Kavli Institute for Particle Astrophysics and Cosmology,
Stanford University, Stanford, CA 94309}

\author{Troy A. Porter}
\affil{
  Santa Cruz Institute for Particle Physics,
  University of California, Santa Cruz, CA 95064
\email{tporter@scipp.ucsc.edu}}

\author{Seth W. Digel\altaffilmark{1}}
\affil{Stanford Linear Accelerator Center,
2575 Sand Hill Road, Menlo Park, CA 94025
\email{digel@stanford.edu}}

\author{Peter F. Michelson\altaffilmark{1}}
\affil{
   Department of Physics,
   Stanford University, Stanford, CA 94305
\email{peterm@stanford.edu}}

\and

\author{Jonathan F. Ormes}
\affil{
Department of Physics and Astronomy,
University of Denver,
Denver, CO 80208   
\email{jormes@du.edu}}

\begin{abstract}
We calculate the \gray{} albedo flux from cosmic-ray (CR)
interactions with the solid rock and ice in Main Belt asteroids
(MBAs),  Jovian and Neptunian Trojan asteroids, and Kuiper Belt
objects (KBOs) using the Moon as a  template.  We show that the
\gray{} albedo for the Main Belt,  Trojans, and Kuiper Belt strongly
depends on the small-body  size distribution of each system.  Based on
an analysis of the  \emph{Energetic Gamma Ray Experiment Telescope}
(EGRET)  data we infer that the diffuse emission from the MBAs,
Trojans, and KBOs has an integrated flux of less than $\sim$$6 \times
10^{-6}$ cm$^{-2}$ s$^{-1}$ (100--500 MeV),  which corresponds to
$\sim$12 times the Lunar albedo, and may be detectable by the
forthcoming \emph{Gamma Ray Large Area Space Telescope} (GLAST).  If
detected by GLAST,  it can provide unique direct information about the
number of small bodies in each system that is difficult to assess by
any other method.  Additionally, the KBO albedo flux can be used to
probe the spectrum of  CR nuclei at close-to-interstellar conditions.
The orbits of  MBAs, Trojans, and KBOs are distributed near the
ecliptic, which passes through the Galactic center and high Galactic
latitudes.  Therefore, the asteroid \gray{} albedo has to be taken
into account when analyzing weak \gray{} sources close to the
ecliptic, especially near the Galactic center and for signals at  high
Galactic latitudes, such as the extragalactic \gray{} emission.  The
asteroid albedo spectrum also exhibits a 511~keV line due to
secondary positrons annihilating in the rock.  This may be an
important and previously unrecognized celestial foreground for the
\emph{INTErnational Gamma-Ray Astrophysics Laboratory} (INTEGRAL)
observations of the Galactic 511~keV line emission including the
direction of the Galactic center.
\end{abstract}


\section{Introduction}\label{intro}
The populations of small solar system bodies (SSSB) in the asteroid
belt  between Mars and Jupiter, Jovian and Neptunian Trojans, and in
the Kuiper Belt beyond Neptune's orbit  (often called also
trans-Neptunian objects -- TNOs) remain the least explored members of
the solar system.  A majority of the MBAs and KBOs have their orbits
distributed  near the ecliptic with a FWHM of the order of 10$^\circ$
in ecliptic latitude \citep{Binzel1999,Brown2001}. The spatial and size 
distributions of these objects provides important information  about
the dynamical evolution of the solar system.  Extending our knowledge
of the size distribution of these objects below current
sub-kilometer size limits of optical
\citep[e.g.,][]{Ivezic2001,Wiegert2007} and infrared
\citep[e.g.,][]{Tedesco2002,Yoshida2003} measurements would provide
additional information on the accretion/collision and depletion
processes that formed the populations of SSSBs\footnote{We note
that \citet{Babich2007} have suggested a method to place constraints upon
the mass, distance, and size distribution of TNOs using spectral
distortions of the CMB.}. 
In this paper we show
that the CR-induced \gray{} albedo of these systems  may be bright
enough to be detected with a \gray{} telescope such as GLAST and/or
INTEGRAL and/or Soft Gamma-ray Detector (SGD) aboard the NeXT satellite 
\citep{Takahashi2006}
(see our estimates below), and can allow us to probe the size
distribution of SSSBs down to a few metres.  Additionally,
the \gray{} emission of these systems may comprise a ``diffuse''
\gray{} foreground that should be taken into account when evaluating the
flux and spectra of \gray{} sources near the ecliptic.  Our
preliminary results are presented in \citet{Moskalenko2008}.

The Galactic center is a region crowded with \gray{} sources and is
one of  the preferred places to look for \gray{} signatures of dark
matter (DM).  An extensive literature on the subject exists, e.g.,
\citet{Bergstrom1998}, \citet{Zaharijas2006}, \citet{Finkbeiner2007},
\citet{Hooper2007}, \citet{Baltz2007}; also references in these
papers.  The ecliptic crosses the Galactic equator near the Galactic
center almost  perpendicularly with inclination $\sim$$86.5^\circ$,
and underestimation of the SSSB albedo foreground may lead to errors
in the analysis of weak or extended sources in this region.

The Galactic center region also harbors the enigmatic source of the
511 keV positron annihilation line observed by the \emph{Oriented
Scintillation  Spectrometer Experiment} (OSSE)
\citep[e.g.,][]{Purcell1997}  and INTEGRAL
\citep[e.g.,][]{Knodlseder2005,Weidenspointner2006}.  The distribution
of the annihilation line  does not match the distribution of any
positron source candidate, e.g., pulsars, supernova remnants,
binaries,  radioactive isotopes, such as $^{26}$Al, etc.  A number of
excellent discussions on the origin of this emission are available in
the literature, ranging from positron focusing by the regular Galactic
magnetic field to DM annihilation  \citep[][and references therein]
{Guessoum2005,Jean2006,Prantzos2006,Finkbeiner2007,Hooper2007}.  Our
calculations (detailed below) indicate the SSSB CR-induced albedo
spectrum  should exhibit a 511 keV line due to secondary positrons
annihilating in the rock.  Since the target material (rock, ice) is
solid,  the line has to be very narrow.  This emission
produces a previously unrecognized celestial foreground to the 511
keV flux including the direction of the Galactic center.

At higher energies, above $\sim$30 MeV, regions at high Galactic
latitudes are conventionally used to derive the level of the
extragalactic \gray{} emission by comparing a model of the diffuse
Galactic emission to the point-source-subtracted skymaps  and
extrapolating to zero model flux
\citep[e.g.,][]{Sreekumar1998,SMR2004b}.  The remainder is assumed to
represent the level of the isotropic, presumably  extragalactic
emission.  However, recent studies have predicted another  important
foreground component with a broad distribution on the sky originating
from the inverse  Compton scattering of solar photons by CR electrons
in the heliosphere \citep{MPD2006,Orlando2007}, which has to be
included in the analysis of the diffuse emission.  A reanalysis of the
EGRET data revealed this broad component, in agreement with the
predictions \citep{Orlando2007b}.  Since the ecliptic  passes through
high Galactic latitudes, the SSSB albedo flux also may need to be
taken into account when  analysing the weak  extragalactic component.

\section{Small solar system bodies}\label{sssb}
The asteroid mass and size distributions are thought to be governed by
collisional evolution and accretion.  Collisions between asteroids
give rise to  a cascade of fragments, shifting mass toward smaller
sizes, while slow accretion leads to the growth of the latter.
The first comprehensive analytical
description of such a collisional cascade is  given by
\citet{Dohnanyi1969}.  Under the assumptions of scaling of the
collisional response parameters and an upper cutoff in mass, the
relaxed size and mass distributions approach power-laws:
\begin{eqnarray}
dN &=& a m^{-k} dm \label{eq1}\\ dN &=& b r^{-n} dr,\label{eq2}
\end{eqnarray}
where $m$ is the asteroid mass, $r$ is the asteroid radius, and
$a,b,k,n$ are constants.  These equilibrium distributions extend over
all size and mass ranges of the population except near its high-mass
end.  The constants in eqs.~(\ref{eq1}), (\ref{eq2}) are not
independent.  If  all asteroids have the same density $\rho$, one
obtains $n=3k-2$ and $b=3a(4\pi\rho/3)^{1-k}$ (see eqs.~[\ref{eq3}],
[\ref{eq4}]).  For a pure \citeauthor{Dohnanyi1969} cascade $k=11/6$
and $n=3.5$.

However, collisional response parameters are not size-independent,
e.g., the energy per unit target mass
delivered by the projectile required for catastrophic disruption
of the target (the so-called critical specific energy)
depends on the radius of the
body, and the single power-laws (eqs.~[\ref{eq1}], [\ref{eq2}])
break.  Even though the sizes of asteroids generally can not be directly
observed (except by a small number of asteroids studied 
by spacecraft flybys, by stellar occultation, or those well observed by radar) 
and are instead estimated using apparent magnitude, optical and 
infrared albedos,
and distances, the information collected on a large sample of
MBAs\footnote{The Minor Planet Center supports a database for all
observed SSSBs: http://www.cfa.harvard.edu/iau/mpc.html} seems to
confirm that the real distribution departs from a single power law, at
least for objects larger than a few kilometers.  Smaller sizes are very
difficult to detect, and one has bear in mind the observational bias
of the incompleteness of the small (dim) asteroid sample.  Though
de-biasing can be attempted \citep[e.g.,][]{Jedicke1998}, a large
ambiguity still remains.

\begin{figure}[t]
\centerline{
\includegraphics[width=3.6in]{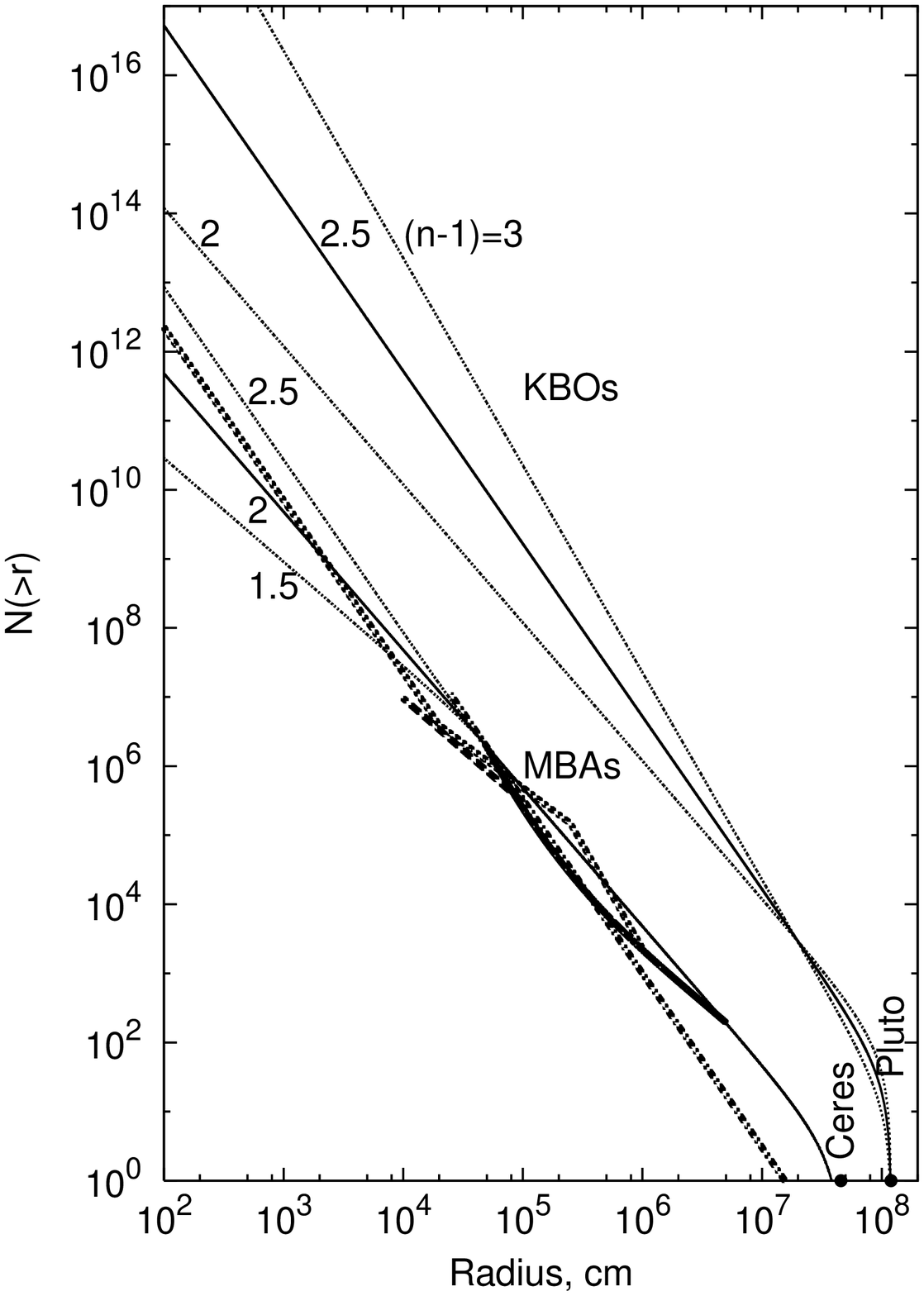}}
\caption{Cumulative size distribution $N(>r)$ of KBOs (upper set of
lines) and MBAs (lower set of lines).  Line coding: thick dash-dot
line -- \citet{Binzel1999}, thick dashes -- \citet{Tedesco2002}, thick
solid -- \citet{Tedesco2005}, thick dots -- parameterization proposed
by \citet{Cheng2004}.  Our parameterizations are shown by thin lines
(solid, dotted) where the numbers show the \emph{cumulative}
power-law index $(n-1)$ of a particular distribution.  Thin solid lines
are our adapted distributions: index 2.0 ($n=3.0$) for MBAs, and 2.5 ($n=3.5$) for KBOs.
Thin dotted lines show the range discussed in the paper. See text for
details.}
\label{fig1a}
\end{figure}

Figure \ref{fig1a} shows the MBA size distributions as published in the
literature and those used in this paper.
For the MBAs
larger than diameter ${\cal D}$ (km),  \citet{Binzel1999} give
$N(>{\cal D})=1.9\times10^6 {\cal D}^{-2.52}$ (the authors do not give
the range of sizes, so we  adopted a cut at ${\cal D}\sim0.5$ km).
\citet{Tedesco2002} give  $\log N(>{\cal
D})=(5.9324\pm0.0016)-(1.5021\pm0.045)\log {\cal D}$ for $0.2 {\rm\
km} < {\cal D} < 2 {\rm\ km}$  based on \emph{Infrared Space
Observatory} (ISO) observations.  Using a sample of more than
$6\times10^4$ MBAs to a limiting magnitude  of $V$$\sim$21,
\citet{Jedicke1998} found a change in the slope of the cumulative
distribution from  --2.25 for 1 km $\la {\cal D} \la $10 km to --4.00
for 10 km $\la {\cal D} \la$ few 10s of km.  Based on observations of
$\sim$13000 MBAs  by the Sloan Digital Sky Survey (SDSS),
\citet{Ivezic2001} found that the  cumulative size distribution
resembles a broken power-law, $\propto {\cal D}^{-2.3}$ for 0.4 km
$\la r \la$ 5 km, and $\propto {\cal D}^{-4}$ for 5 km  $\la r \la$ 40
km, and is independent of the heliospheric distance.  Finally,
\citet{Tedesco2005} gives a fit to data between 1 km $\la {\cal D}
\la$ 100 km, $\log N(>{\cal D})=6.275\pm0.013-(3.214\pm0.056)\log{\cal
D} +(0.974\pm0.066)\log^2{\cal D}-(0.182\pm0.022)\log^3{\cal D}$, but
extrapolation to smaller sizes is invalid.  The size distribution
below  $\sim$1 km is essentially unexplored territory.  One piece of
evidence comes from the size distribution of ejecta blocks on 433
Eros.  Based on the block distribution over a size range 0.1 -- 150 m,
\citet{Cheng2004} argued that these data support a cumulative index 2.5
extrapolation down to sizes $\sim$1 m.  Our distribution with a single
cumulative
index $(n-1)=2$ (thin solid line), detailed in the next Section, seems
to match the global size distribution determined from various types of
observations in the wide range of radii $10^2 - 10^7$ cm.  We will use
this  distribution in our estimates of the MBA albedo, below.

The dynamical estimate of the total mass of the asteroid belt is about
$(3.6\pm0.4)\times10^{24}$ g \citep{Krasinsky2002} or close to 5\% of
the mass of the Moon.   The total mass is dominated by large bodies, while
the \gray{} albedo is dominated by very small bodies.  The largest
MBA, 1 Ceres, comprises about 30\% of the total mass of the asteroid
belt alone.  However, it does not provide a  restriction on the size
(and mass) distribution of small bodies.  Current estimates indicate
the total number of MBAs above 1 km in diameter  is
$(1.2-1.9)\times10^6$ \citep{Binzel1999,Tedesco2002,Tedesco2005}.  Our
adopted distribution with $n=3$ gives a number near the upper end of
this range,  $1.92\times10^6$, while also putting  the total number of
MBAs with $r>1$ m  at $\sim5\times10^{11}$ (Figure \ref{fig1a}).  To
get an idea of how the MBA albedo flux depends on the extrapolation to
small radii ${\cal D}<1$ km, we also consider broken power-law
distributions with indices 2.5 and 3.5 below 1 km in diameter,
retaining an index 3 for sizes larger than 1 km.

The densities of most MBAs lie in the range 1.0 -- 3.5 g cm$^{-3}$ 
\citep{Binzel1999} while the densities of particular asteroid classes
can vary broadly, 
1.23 -- 1.40 g cm$^{-3}$ for carbonaceous, 
2.65 -- 2.75 g cm$^{-3}$ for silicate, and 
4.75 -- 5.82 g cm$^{-3}$ for metallic bodies \citep{Krasinsky2002}.
We adopt an average density $\rho=2$ g cm$^{-3}$.

Most MBAs have a semimajor axis between 2.1 and 3.3 AU with a low
eccenticity orbit.  In our estimates we assume an average circular
orbit with radius  $\ell\sim2.7$ AU.

The Jovian Trojan populations of asteroids are collections of bodies in
the same orbit as Jupiter (semimajor axis $\ell\sim5.2$ AU) located at
the $L_4$ and $L_5$ Lagrange points of the Jupiter-Sun  system.  The
Trojans are thus  concentrated in two regions rather than distributed
over the entire  ecliptic as for the MBAs.  The total mass of the
Jovian Trojans is estimated to be  $\sim10^{-4}$ $M_\oplus$ where
$M_\oplus$ is the mass of the  Earth with a differential power-law
index $n \simeq 3$  in the size range 2 km to 20 km
\citep{Jewitt2000,Yoshida2005}, similar to  MBAs, giving a number of
objects $\geq$ 1 km in diameter $\sim 1.3 \times 10^6$
\citep{Jewitt2000}.  The combined mass of these objects is 
approximately the same as for the MBAs.
The number of objects $\geq$ 1 km in diameter and the
power-law index $n \simeq 3$ makes their size distribution very
similar to that of MBAs.

The mass density of SSSBs in this group varies significantly:
estimates for the binary Trojan 617 Patroclus are less
than water  ice $\rho=0.8^{+0.2} _{-0.1}$ g cm$^{-3}$
\citep{Marchis2006}, as are those  for other Trojan binaries
$\rho\sim0.6-0.8$ g cm$^{-3}$ \citep{Mann2007},  while 624 Hektor is
somewhat denser $\rho=2.48^{+0.292}_{-0.080}$ g cm$^{-3}$
\citep{Lacerda2007}.  In our calculations we adopt an average density
$\rho=1$ g cm$^{-3}$ as a  compromise between these bounds.

We also consider icy bodies and comets in the Kuiper Belt \citep[for
a review see][]{Luu2002} and the conjoining  innermost part of the
Oort  Cloud\footnote{The Oort Cloud of comets
\citep[e.g.,][]{Stern2003}  is thought to occupy a vast space between
50 and 50000 AU from the Sun and  also contributes to the celestial
\gray{} foreground. However, its exact  mass and distribution are
poorly constrained. We are planning to investigate  limits on the
albedo of the Oort Cloud in a forthcoming paper.},  but call them all
KBOs for simplicity.  The KBOs are not uniformly distributed, with at
least three dynamically  distinct populations identified: the
Classical Disk, the Scattered Disk  with large eccentricities and
inclinations, and ``Plutinos'' around the 3:2  mean motion resonance
with Neptune at 39.4 AU.  Kuiper Belt Objects are distributed between
30 -- 100 AU  \citep[][and references therein]{Backman1995} with
surface number density $\sigma(\ell)=A \ell^{-\alpha}$
\citep{Jewitt1995,Backman1995},  where $A$ is a constant determined in
eq.~(\ref{eq11}), and $\alpha=2$.  The total mass is  estimated to be
in the range $\sim$0.01--0.3 $M_\oplus$,
while the most often used  value is $\sim$0.1 $M_\oplus$
\citep{Luu2002}.  The density of small icy bodies and comets is
$\sim$0.5 g cm$^{-3}$  \citep{Asphaug1994,Solem1994}.

The KBO size distribution is much more difficult to determine because
of their dimness.  It is widely believed that the TNOs are dynamically
related to the Centaurs (planetesimals distributed between Jupiter 
and Neptune that are 
in crossing orbits of the giant gas planets), and to
the Jupiter-family group of ecliptic comets that may be  objects that
were knocked inwards from the Kuiper belt.  The KBO size distribution
is determined by very indirect methods such as measuring the sizes of
the nuclei of the ecliptic comets  (and making assumptions on how they
evolve during their repeated passages  through the inner solar
system), and Centaurs,  as well as impact craters on the Galilean
satellites of Jupiter.  The estimates of the size distribution for the
cometary nuclei range from $n=2.6$ to 3.7 in the range $r\sim 1-10$
km, for KBOs -- 3.7--4.45 ($r>20$ km), and for Centaurs -- 3.7--4.0
(an appropriate discussion can be found, e.g., in
\citealt{Bernstein2004}, \citealt{Toth2005}, \citealt{Tancredi2006},
and references therein).  If the index is $\geq 4.0$, the mass of the
total population is dominated by the smallest bodies.  However, there
are some reasons to believe that the size distribution begins to
flatten well above 1 km in size.  Collisional  evolution simulations
\citep{Kenyon2004} show that the size distribution is a power law with
index $\sim$4.5 for large bodies ($r\ga10-100$ km) and $\sim3.5-4$ for
small bodies ($r\la0.1-1$ km) for a wide range of bulk properties,
initial masses, and orbital parameters.  Adopting a conservative value
of $n=3.5$, we obtain the total number of comets (${\cal D}>1$ km) at
$\sim9\times10^9$, which is in agreement with other estimates
\citep[e.g.,][]{Stern2003}.


There are also large populations of Centaurs $N({\cal D}>1\ {\rm
km})\sim10^8$  \citep{Sheppard2005} between the orbits of Jupiter and
Neptune, and Neptunian Trojans at the $L_4$ and $L_5$ points of the Neptune-Sun
system. 
The number of large Neptunian Trojans
(${\cal D} > 80$ km) outnumbers the number of large Jovian Trojans by a
factor $\sim 10$ \citep{Sheppard2006}.
Their power-law index may be close to
that of the KBOs $n\sim3.5$  making their \gray{} albedo essentially
brighter  than MBAs and Jovian Trojans at the same distance.  While
Centaurs are scattered between Jupiter and Neptune, the positions of
Neptunian Trojans are well known so that the detection of  a \gray{}
albedo signal may be simplified.

\placefigure{f1.ps}

\section{Calculations}

We use the Lunar albedo spectrum as an approximation of the SSSB
albedo for two main reasons: (i) the Moon is a solid body in which the
CR cascade in the rock develops similarly, and (ii) its proximity to
the Earth allows it to be easily detectable by \gray{} telescopes. The
spectrum of \gray{s} from the Moon has been calculated  recently
\citep{MP2007a,MP2007b} using the \geant\ Monte Carlo framework to
simulate the CR cascade development in a Lunar rock target (regolith).
It has been shown that the Lunar albedo spectrum is very steep with
an  effective cutoff around 3--4 GeV in agreement with observations
\citep{Thompson1997}.  The central part of the disk of the Moon has an
even steeper  spectrum with an effective cutoff at $\sim$600 MeV.  The
emission above $\sim$10 MeV is thus dominated by the thin rim where CR
particles interact close-to tangentially with the surface and the
high-energy secondary \grays{} shower out of the Moon toward the
observer.  In contrast to other astrophysical sources, the albedo
spectrum of the Moon is well understood, including its absolute
normalization; this makes it a useful template for estimations of the
CR-induced albedo of SSSBs without an atmosphere. 
Since the
Moon functions as a standard (\gray{}) candle, 
in our estimates we use the flux of the Moon as our standard and introduce 
the term ``Lunar albedo flux units.''

If the SSSB size distribution $dN/dr$ is known, it can be directly
substituted into eq.~(\ref{eq6}) to estimate the \gray{} albedo flux.  Below, 
we derive this albedo flux assuming that the size distribution for a 
single SSSB population is a
simple power law where the normalization has to be obtained from the
total mass of the system.

Let the SSSB mass distribution have the form given by eq.~(\ref{eq1}), 
which can be rewritten as a size distribution (cf.\ eq.~[\ref{eq2}])
\begin{equation}
\frac{dN}{dr} = \frac{dN}{dm}\frac{dm}{dr} = 4\pi\rho r^2 \frac{dN}{dm},
\label{eq3}
\end{equation}

\noindent
where $\rho$ is the average density of the SSSB target. 
Assuming that all SSSBs have a spherical shape, $m = (4\pi/3) \rho r^3$, we get 
%
\begin{equation}
\frac{dN}{dr} = 3 a \left(\frac{4\pi}{3} \rho\right)^{1-k} r^{2-3k}.
\label{eq4}
\end{equation}

\noindent
The normalization for $a$ is obtained from
\begin{equation}
\int_{m_0} ^{m_1} m \frac{dN}{dm} dm = f M_{\moon},
\end{equation}

\noindent
where $m_0$ and $m_1$ are the lower and upper SSSB masses, respectively, and
$f M_{\moon}$ is the total mass of the SSSB emitting population considered
as a fraction $f$ of the Moon's mass ($f=0.05$ for MBAs, 
$f=0.1 M_\oplus/M_{\moon}\approx8.13$ for KBOs).
The flux of \gray{s} from such an ensemble of bodies with
size distribution $dN/dr$ is then
\begin{eqnarray}
F & = & F_{\moon} \left( \frac{D_{\moon}}{d} \right)^2
\int_{r_0} ^{r_1} \frac{dN}{dr} \frac{r}{R_{\moon}} dr \label{eq6}\\
& = &  \frac{F_{\moon}}{R_{\moon}} \left( \frac{D_{\moon}}{d} \right)^2
\left(\frac{4\pi}{3}\rho\right)^{1-k} 3 a \int_{r_0} ^{r_1} r^{3 - 3k} dr \nonumber
\end{eqnarray}

\noindent
where
\begin{eqnarray}
&& \int_{r_0} ^{r_1} r^{3 - 3k} dr= \left\{
\begin{array}{ll}
\displaystyle \frac{1}{4-3k}\left(r_1 ^{4-3k} - r_0 ^{4 -3k}\right), & k \neq 4/3 \\
\displaystyle \ln(r_1/r_0), & k = 4/3
\end{array} \right.\nonumber \\
&& a = f M_{\moon} \left\{
\begin{array}{ll}
\displaystyle \left(\frac{4\pi}{3}\rho\right)^{k - 2}\frac{2-k}{r_1^{6-3k} - r_0^{6-3k}}, 
& k \neq 2 \\
\displaystyle \frac{1}{3}\ln^{-1}(r_1/r_0), & k = 2
\end{array} \right. \nonumber 
\end{eqnarray}

\noindent
Here $F_{\moon}$ is the Lunar rim albedo flux for the same incident spectrum of CR particles,
$R_{\moon}= 1.7382\times10^8$ cm is the Lunar radius,
$D_{\moon} \simeq 0.0025$ AU is the Earth-Moon distance, $d$ is the distance (in AU) of
the SSSB population from Earth, and $r_0$ and $r_1$ are the sizes corresponding to
the masses $m_0$ and $m_1$. 

\begin{figure}[t]
\centerline{
\includegraphics[width=3.4in]{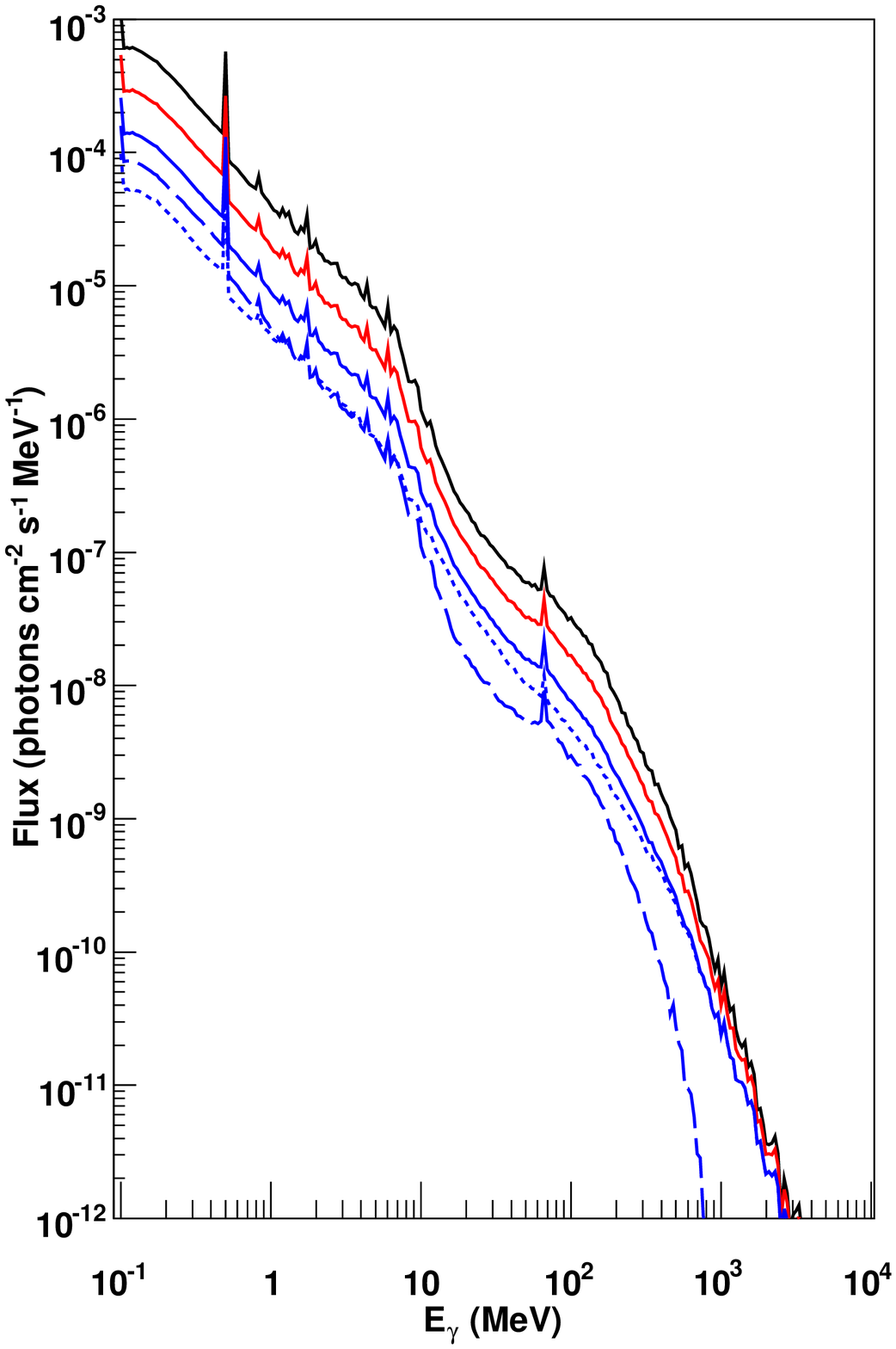}}
\caption{Calculated \gray{} albedo spectrum for CR nuclei interactions
in the Moon rock \cite{MP2007b} for selected modulation potentials.
Line colouring: black, no modulation; red, $\Phi = 500$ MV; blue,
$\Phi = 1500$ MV. Dashed and dotted lines show the albedo of the disk
and the rim correspondingly for the case of $\Phi=1500$ MV.  }
\label{fig:moon1}
\end{figure}

The factor $r/R_{\moon}$ in eq.~(\ref{eq6})
comes from the fact that the albedo of SSSBs is dominated by the emission 
from the rim.
The rim and the disk albedo fluxes of the Moon are about equal at low energies 
(Figure~\ref{fig:moon1}),
with the rim albedo flux considerably dominating above 10 MeV \citep{MP2007b}.
Since the rim albedo flux scales $\propto r$, 
and the inner part of the disk $\propto r^2$, 
as the size of the emitting body decreases it is the rim which 
produces most of the albedo photons for SSSBs.

Assuming $k \neq 4/3, 2$, after some rearrangement we obtain
\begin{equation}
F(\ell,\theta) = \frac{3}{2\pi} f F_{\moon} R_{\moon}^2 \frac{\rho_{\moon}}{\rho}
\left[\frac{D_{\moon}}{d(\ell,\theta)}\right]^2 G(r_1, r_0; k),
\label{eq7}
\end{equation}
\begin{equation}
G(r_1, r_0; k) = \frac{1}{r_1^2}\left[\frac{2-k}{4-3k}\right]
\left[\frac{1 - (r_0/r_1) ^{4 -3k}}{1 - (r_0/r_1)^{6-3k}}\right],
\label{eq8}
\end{equation}

\noindent
where $\rho_{\moon}=3.3$ g cm$^{-3}$ is the mean density of the Moon,
\begin{equation}
d(\ell,\theta)=\cos\theta +\left(\ell^2 -\sin^2\theta\right)^{1/2},
\label{eq9}
\end{equation}
$\ell$ (AU) is the radius of the orbit of the SSSB population 
(for MBAs, $\ell = 2.7$ AU; for KBOs, see below), $\theta$ is the angle
between the line of sight (in the ecliptic) and the direction to the Sun,
and we divided by $2\pi$ to obtain the flux per radian.
The total flux integrated over $\theta$ is
\begin{equation}
F_{\rm tot}=\int_0^{2\pi} F(\ell,\theta) d\theta.
\label{eq10}
\end{equation}

\noindent
For the case of the KBOs distributed between $\ell=30$ and 100 AU, 
an additional integration over $\ell$ is required
\begin{equation}
F_{\rm tot}^{\rm K}
= \int_0^{2\pi}  d\theta 
\int_{30}^{100} F(\ell,\theta) \sigma(\ell) \ell d\ell,
\label{eq11}
\end{equation}

\noindent
where $\sigma(\ell)=A \ell^{-2}$ and $A=-f/\ln0.3$.
These formulae provide fluxes integrated over ecliptic latitude.

For the case of MBAs, our adopted differential size distribution is a 
broken power law
with index $n_1=3k_1-2$ for $r>r_b$ and $n_2=3k_2-2$ for $r<r_b$:

\begin{eqnarray}
  \frac{dN}{dr} &=& 3 f M_{\moon}\left(\frac{4\pi}{3}\rho\right)^{-1}\\
  &\times& \frac{2-k_1}{r_1^{6-3k_1} - r_0^{6-3k_1}}
  \left\{
  \begin{array}{ll}
    \displaystyle r  ^{2-3k_1},                                   & r \geq r_b\\
    \displaystyle r_b^{2-3k_1}\left(\frac{r}{r_b}\right)^{2-3k_2},& r < r_b
  \end{array} \right. \nonumber
\end{eqnarray}
where $r_b=0.5\times10^5$ cm, and we assume $2<n_{1,2}<4$.
This can be inserted into eq.~(\ref{eq6}) to derive corresponding expressions
for the flux from such a population of SSSBs.


We can see that the observed albedo flux gives direct information on
the integral $\int dr\, r\, (dN/dr)$, eq.~(\ref{eq6}),  which can be
used to constrain the effective average radius of the emitting bodies
$\langle r\rangle$ and their total number in the system.
Additionally,  if the size distribution is a single power law, the
observed albedo flux can provide us with information about the
power-law index.  
As can be seen from eq.~(\ref{eq8}),
the function $G(r_1, r_0; k)$ is a
steep function of $k$.  For $k<4/3$, the expression in the last square
brackets (eq.~[\ref{eq8}]) is $\sim$1 since $r_1\gg r_0$ and
$G\approx r_1^{-2} (2-k)/(4-3k)$.  For $k>2$, eq.~(\ref{eq8}) becomes
$G\approx r_0^{-2} (2-k)/(4-3k)$.  In the intermediate region
$4/3<k<2$, $G$ quickly increases with $k$.  For the distribution of
radii we consider the corresponding range for the size distribution index is 
$n=2.5-3.5$ for MBAs and 3.0--4.0 for
KBOs.  This translates into an index, $k$,  for the mass distribution
in the range 1.5--2.0.  In this range $G(r_1, r_0; k)$ changes by 3 --
5 orders of magnitude depending mostly on the assumed value of $r_0$.
This allows a determination of $k$ assuming the  average density of
the asteroid rocks  is  known.  The function $G$ also contains a
dependence on $r_1$, the radius of the largest body.  For the MBAs, we
use Ceres, $r_1=4.565\times10^7$ cm, and for the KBOs we use 134340
Pluto, $r_1=1.195\times10^8$ cm,  but the exact value of $r_1$ does
not change the size distribution significantly and does  not affect
our conclusions.  Figure~\ref{fig1a} shows the adopted size
distributions of MBAs and KBOs which agree well with those given in
the literature.

The question of where most of the heliospheric modulation occurs is
important for the determination of the CR flux at an arbitrary
distance  from the Sun.  The recent crossing of the heliospheric
termination shock by the Voyager 1  spacecraft at  $\sim$94 AU
\citep{Stone2005}, currently at $\sim$104 AU, while Voyager 2 is
still  inside the termination shock, allows unique studies of the
spectra  of CR particles on both sides of the shock.  Low-energy CR
detectors on board the spacecraft indicate that the particle spectra
are significantly different, supporting the conclusion that a
considerable modulation of the CRs occurs near the termination shock.
On the other hand, most of the albedo emission discussed in this paper
is produced by CR particles with  energies  $>$1 GeV; their flux does
not change significantly from local interstellar space down to
$\sim$40 AU, as indicated by current heliospheric models
\citep[e.g.,][]{Langner2006}.  

The Lunar albedo flux, $F_{\moon}$, 
is calculated using the procedure described in \citet{MP2007b}.
To calculate the Lunar albedo at an arbitrary modulation level, we use
the local interstellar (LIS) spectra of CR protons, helium, and positrons, 
as fitted to the 
numerical results of the GALPROP propagation model \citep[][Table 1, 
reacceleration and plain diffusion models]{Ptuskin2006}
as described in \citet[][eq.~(2)
with parameters listed in Table 1]{MP2007b}\footnote{The parameterization 
constants for CR positrons, not given in Table 1 of \citet{MP2007b}, are: 
$J_0 =44.8143$, $a_1 = 1$, $b_1 = 0.594634$, $c_1 = -9.14888$, 
$a_2 = -605.291$, $b_2 = 1.53611$, $c_2 = -7.27809$, 
$a_3 = 1.18135$, $b_3 = 0.365787$, $c_3 = -3.51576$.}.
The CR particle flux at an arbitrary phase of solar activity 
can then be estimated using the force-field approximation \citep{Gleeson1968}:
\begin{equation}
\frac{dJ_p(E_k)}{dE_k}=\frac{dJ_p^\infty(E_k+\Phi Z/A)}{dE_k}\frac{E^2-M^2}{(E+\Phi Z/A)^2-M^2},
\end{equation}
where $dJ_p^\infty/dE_k$ is the LIS spectrum of the CR species, $E_k$ is the
kinetic energy per nucleon, $E$ is the total energy per nucleon,
$\Phi$ is the modulation potential, $Z$ and $A$ are the nuclear charge
and atomic number correspondingly, and $M$ is the nucleon mass.  
The modulation potential $\Phi(\ell)$ at an arbitrary distance $\ell$ 
from the Sun
can be 
calculated using the expressions derived in \citet{MPD2006}, their
eqs.~(7), (8).

\placefigure{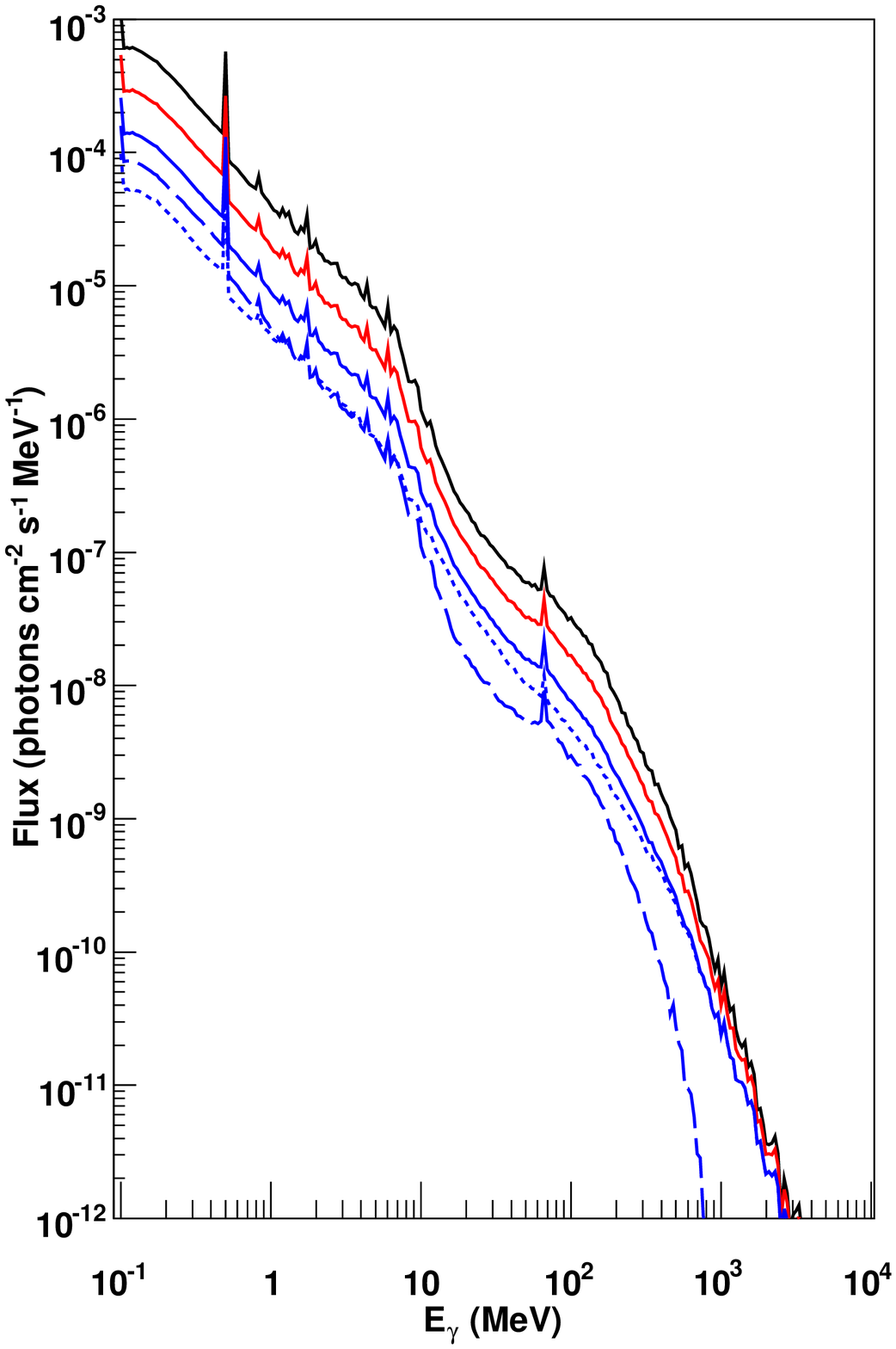}
\placefigure{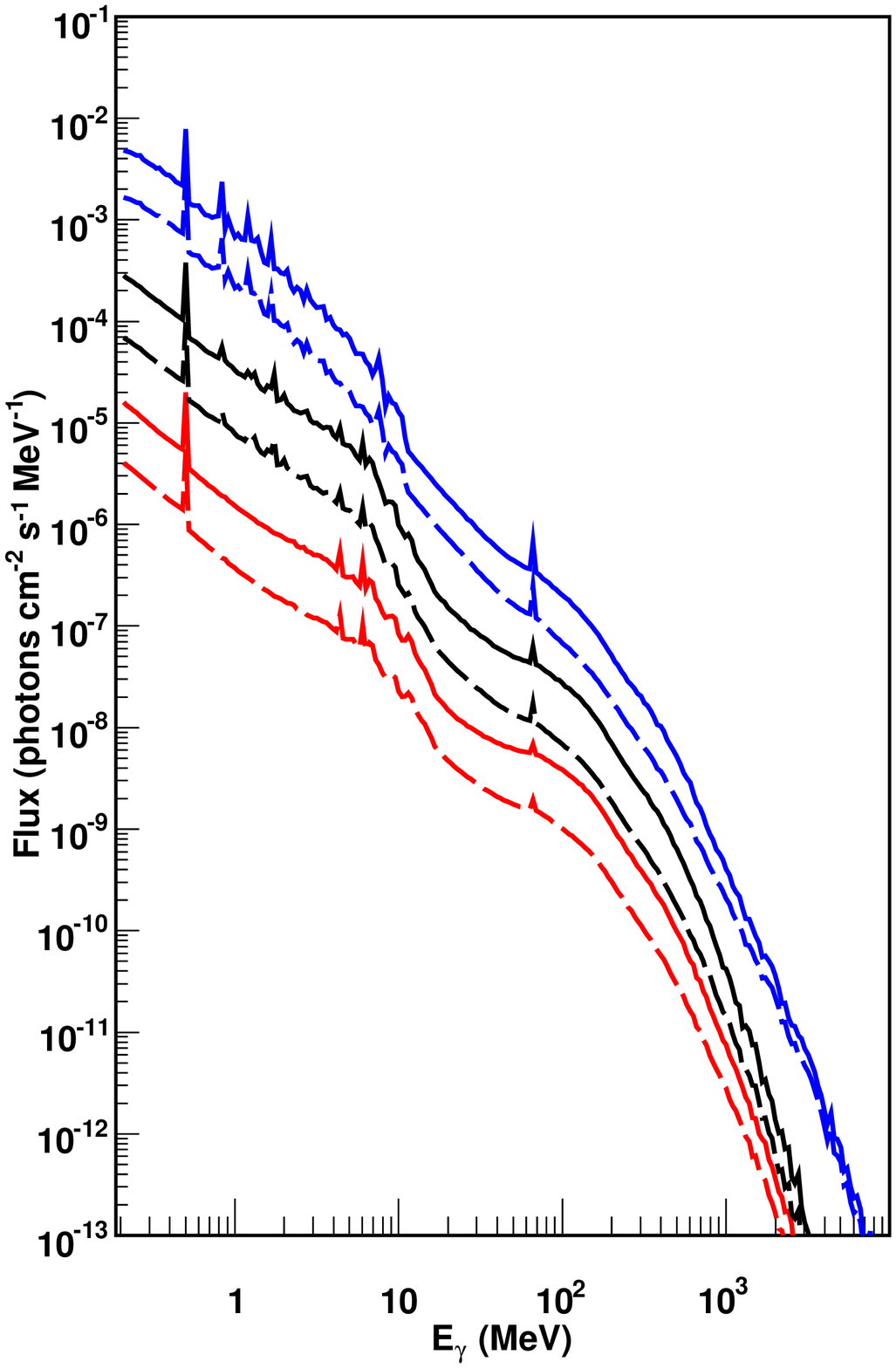}
\placefigure{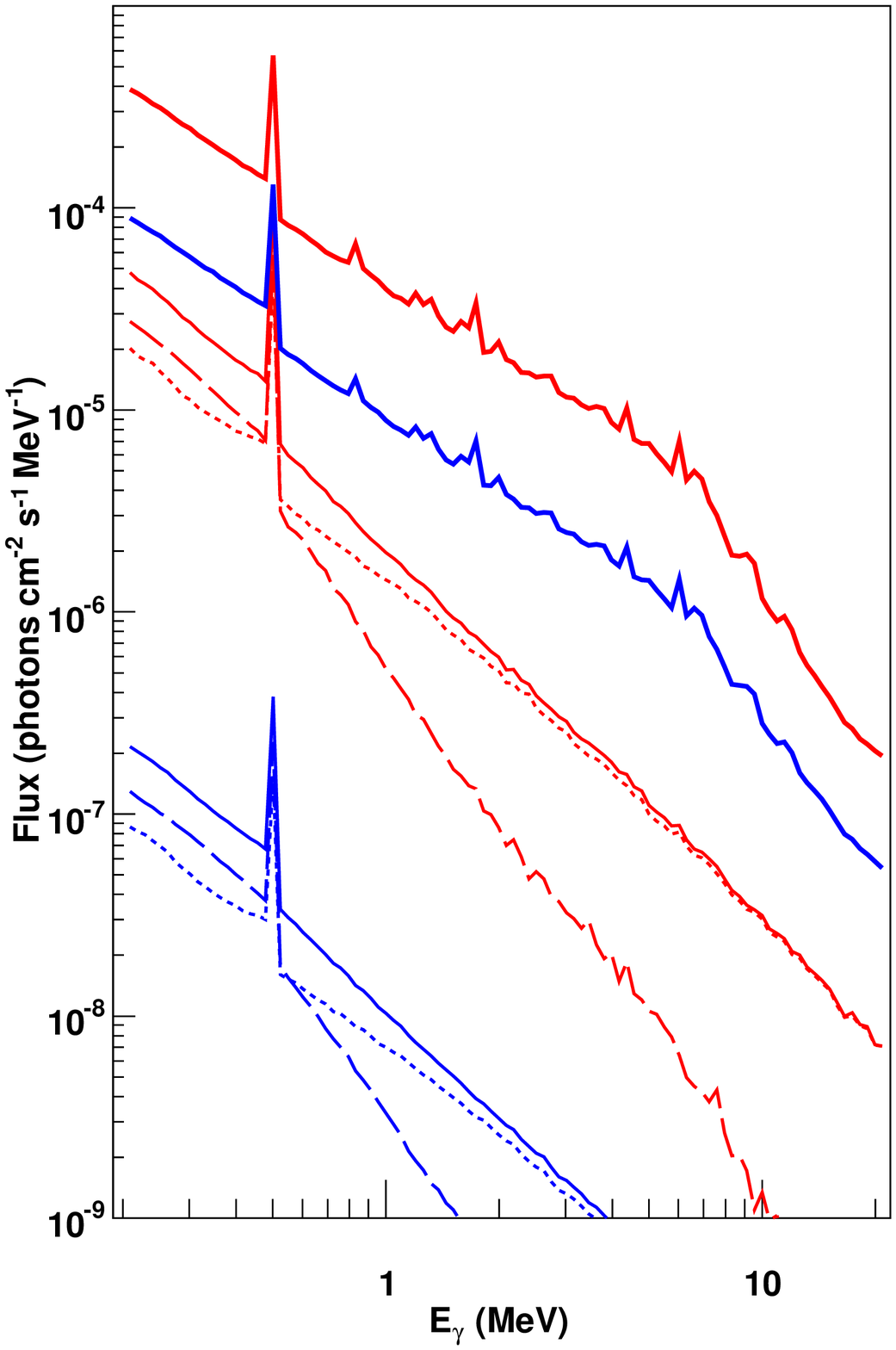}

\begin{figure}[t]
\centerline{
\includegraphics[width=3.4in]{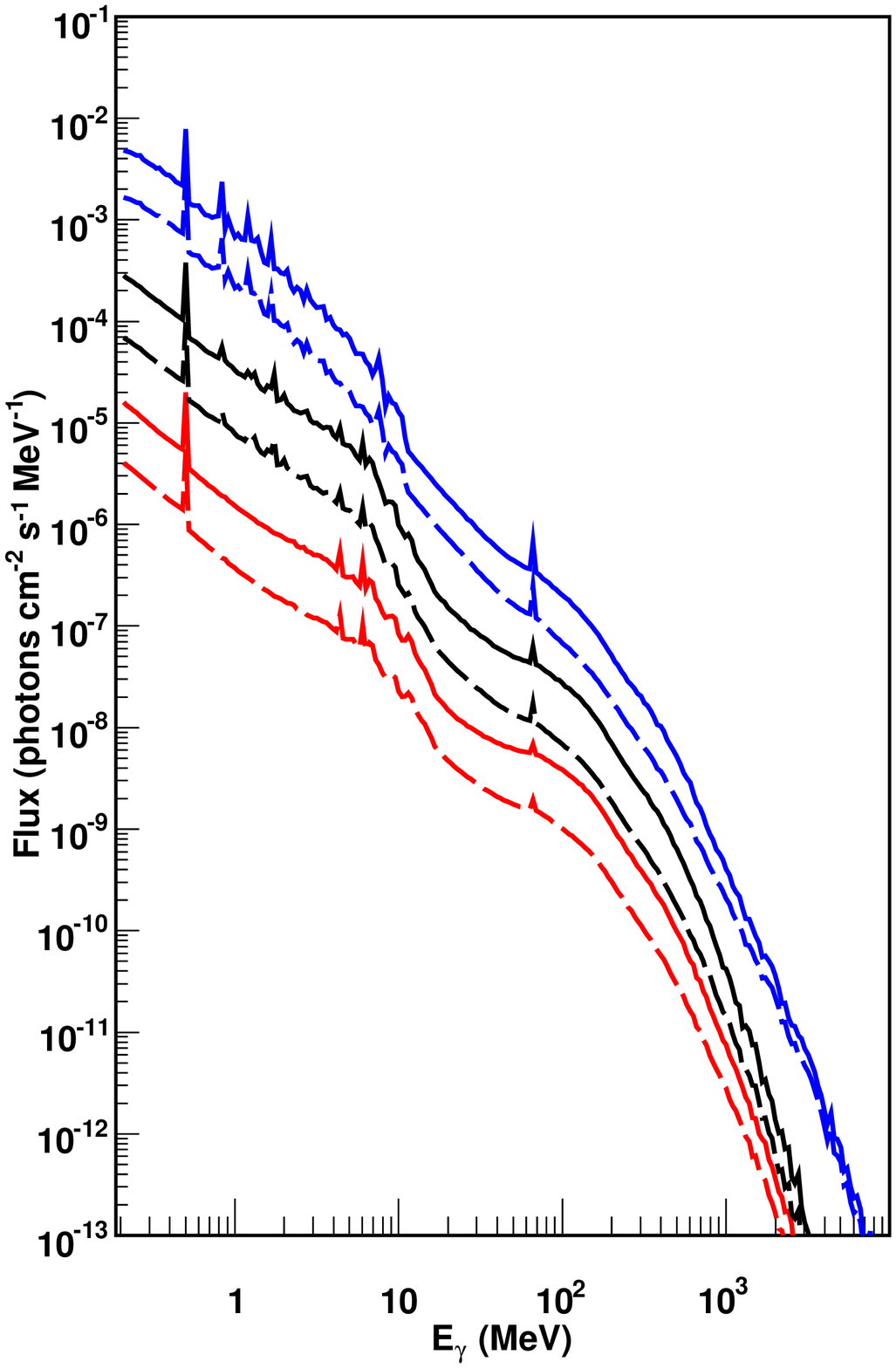}}
\caption{Calculated \gray{} albedo spectrum of a Moon-sized body at the
Lunar distance composed of moon rock (black), iron ($\times10$, blue), or water ice ($\times0.1$, red).
Line-styles: solid, no modulation; long-dashed, $\Phi = 1500$ MV.}
\label{fig:moon}
\end{figure}

\begin{figure}[t]
\centerline{
\includegraphics[width=3.4in]{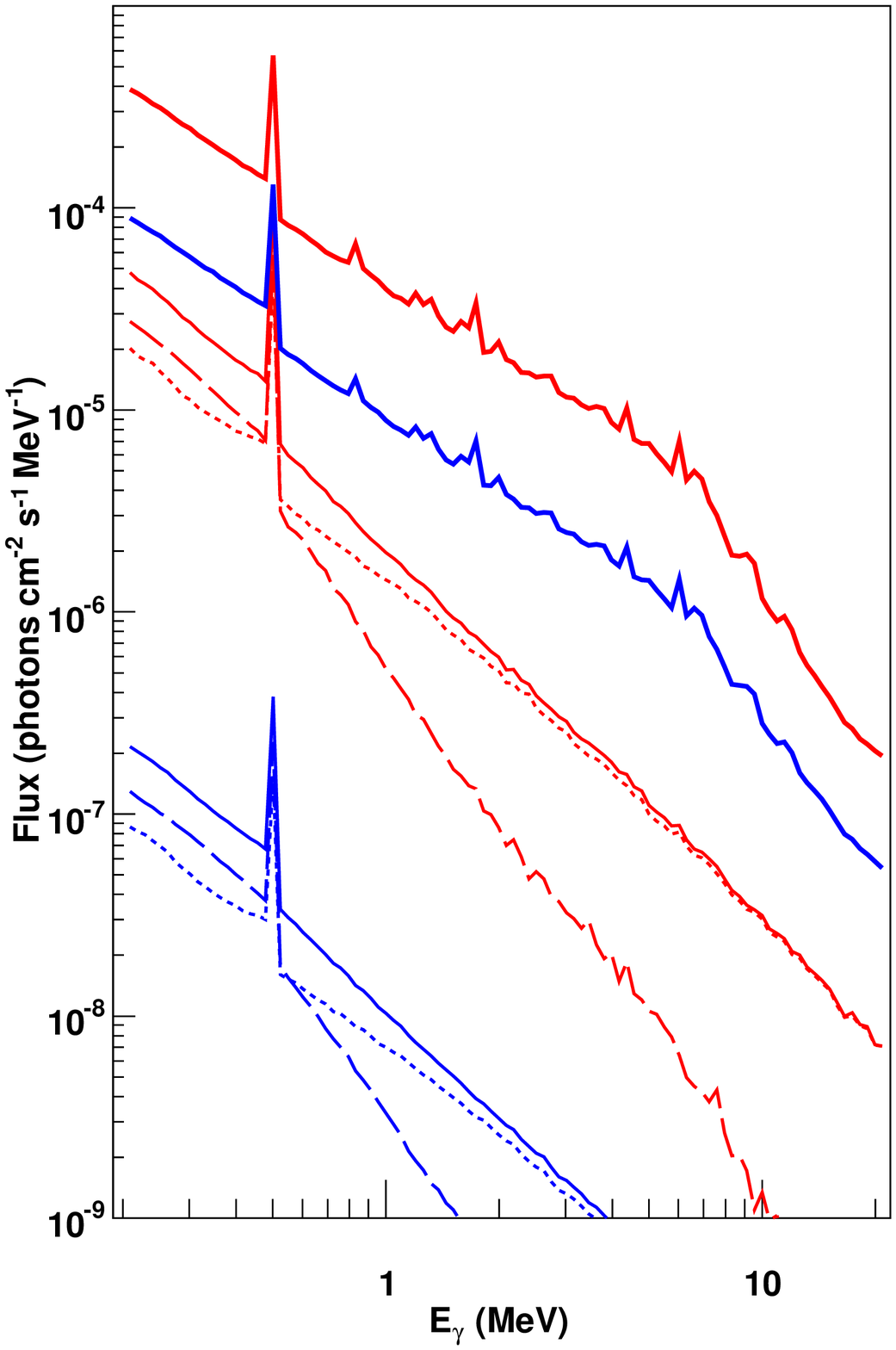}}
\caption{Calculated \gray{} albedo spectrum showing components below 20 MeV for
no modulation (red) and modulation level 1500 MV (blue). Line-styles:
long dash: positron induced \gray{s} from center; short dash:
CR positron induced \gray{s} from rim; thin solid: total CR positron
induced \gray{s}; thick solid: total \gray{} emission from CR positrons
and nucleons.
}
\label{fig:moon2}
\end{figure}

Figure~\ref{fig:moon1} shows the Lunar albedo spectrum for different
modulation potentials $\Phi=0, 500, 1500$ MV.
The no modulation case ($\Phi=0$) corresponds to the upper limit of the 
KBO albedo, with moderate modulation ($\Phi=500$ MV) corresponding to
the MBA albedo. The difference in brightness below $\sim$1 GeV
due to the incident CR flux only (no modulation vs.\ moderate modulation)
is as large as a factor of $\sim$2--3. Also shown are the 
components of the albedo spectrum (center, rim) for $\Phi=1500$ MV.

Figure~\ref{fig:moon} shows the albedo spectrum of the Moon as if the
Lunar surface was made  of different materials: water ice (multiplied
by a factor of 0.1),  regolith, and iron (multiplied by a factor of
10).  The  plot shows the albedo spectra for two limiting cases, no
solar modulation ($\Phi=0$) and solar maximum conditions at 1 AU
($\Phi=1500$ MV).  The low-energy parts of the spectra ($<$10 MeV)
from different materials are considerably different and the nuclear
emission lines can be used to distinguish between the materials.  The
high-energy parts are essentially featureless and have similar shape.
The flux between iron and water ice changes by a factor of $\sim$2
above 100 MeV with the latter producing the larger flux.  Above
$\sim$100 MeV the  regolith albedo approaches the water ice albedo.

The 511 keV line in  Figures~\ref{fig:moon1} and \ref{fig:moon} is due
to the annihilation of positrons produced by CR cascades in the  solid
target (iron, regolith, ice).  In Figure~\ref{fig:moon1}, the albedo
spectrum also includes the contribution by CR positrons in the Lunar
rock target (see below).  Since the rock is solid, secondary positrons
quickly thermalize and produce a narrow annihilation line.  Its width
is determined by the energy bin size adopted in the simulation.

Figure~\ref{fig:moon2} shows the components of the albedo spectrum
(Figure~\ref{fig:moon1}) below 20 MeV.  The thick solid lines are the
total albedo flux due to the CR proton, helium, and positron
interactions with regolith for  no modulation (upper, red) and
modulation level 1500 MV (lower, blue).  The thin solid lines show the
albedo spectrum due to CR positron interactions with regolith for the
same cases of no modulation (upper, red) and modulation  level 1500 MV
(lower, blue).  The dashed and dotted lines show the components of the
CR positron induced \gray{s}, from the center and the rim,
correspondingly.

\placefigure{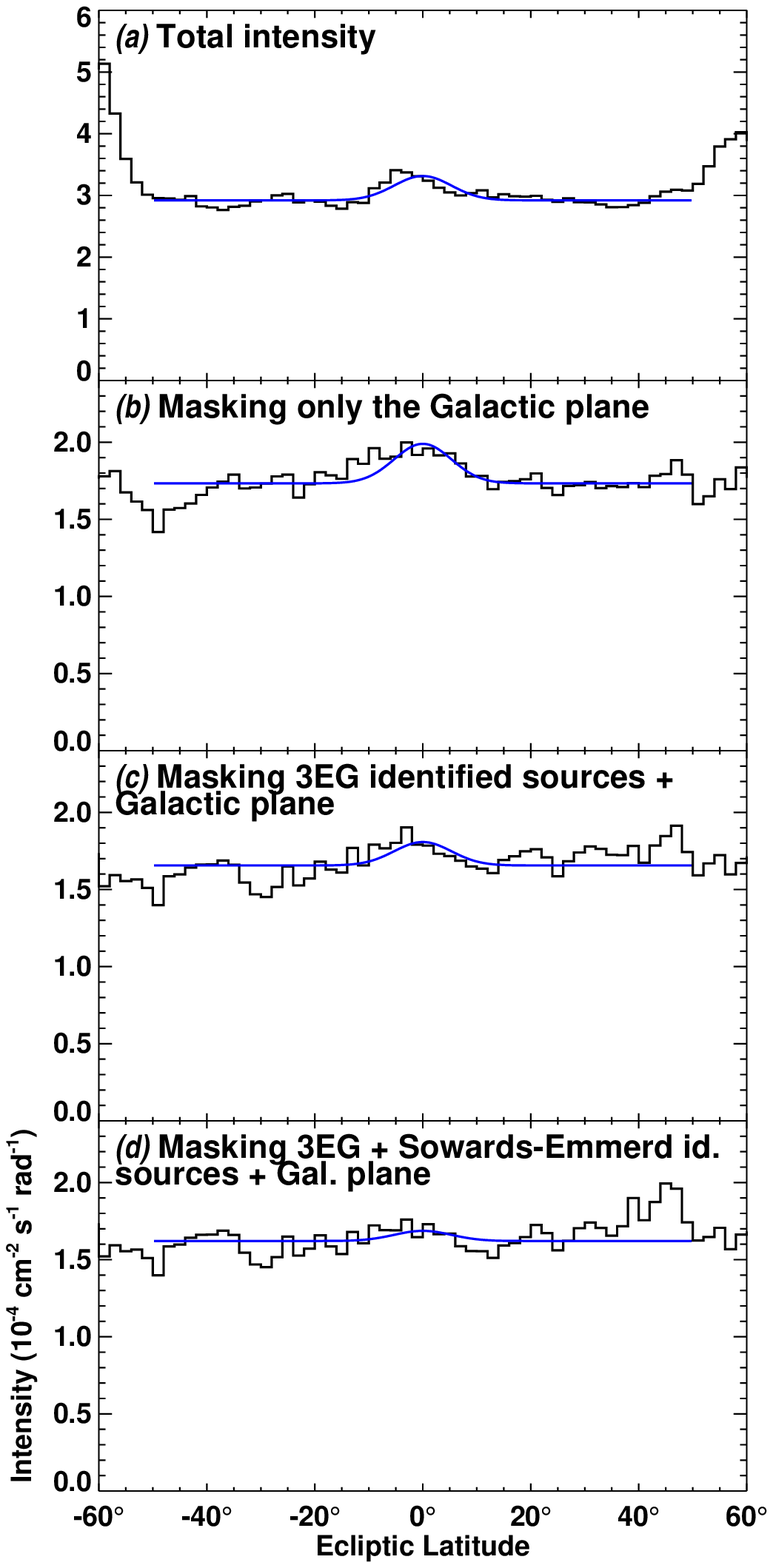}

\begin{figure}[t]
\centerline{
\includegraphics[width=3in]{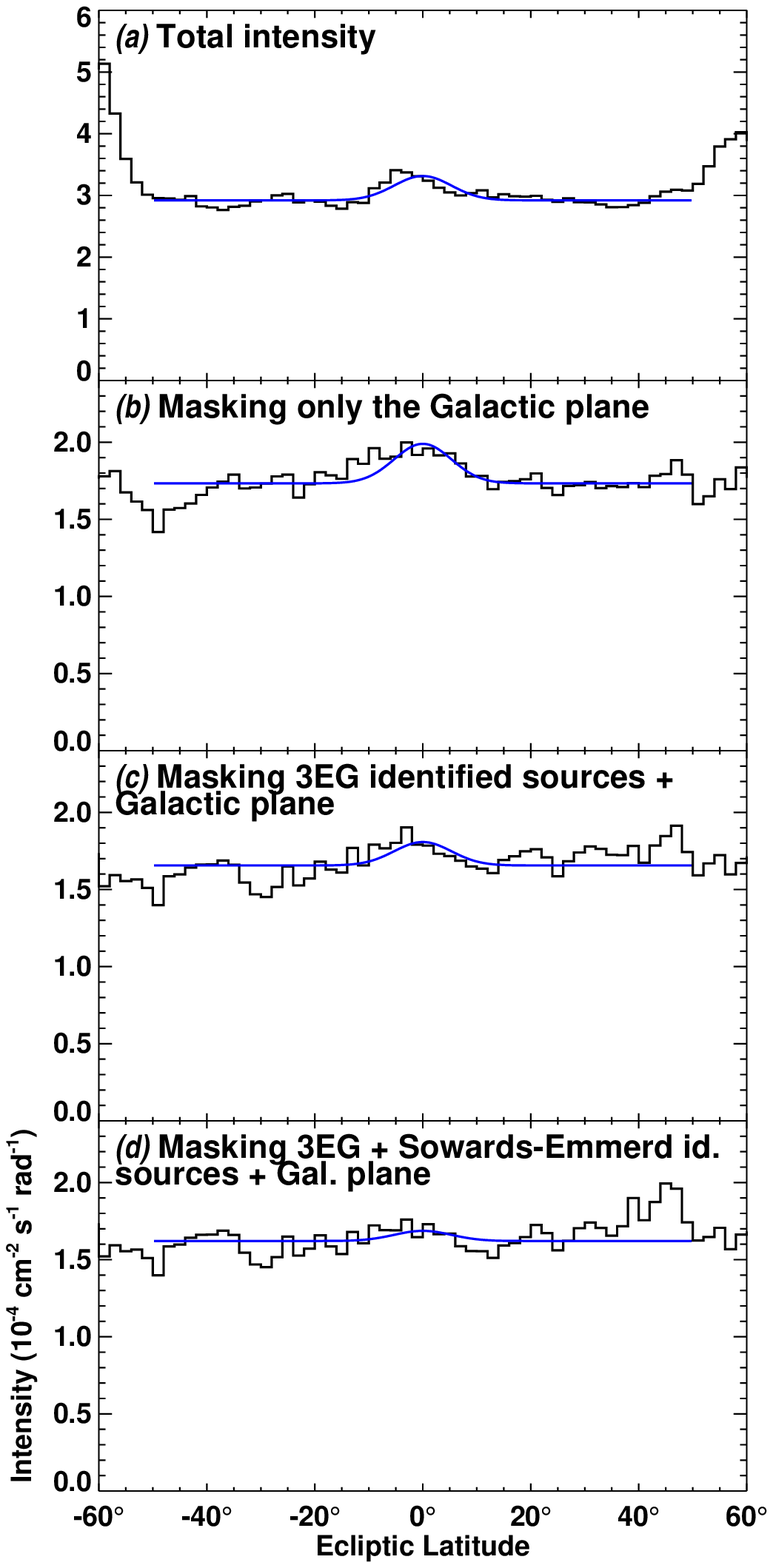}}
\caption{
Profiles of \gray{} intensity with $\beta$ derived from EGRET data
as described in the text.  The energy range is 100--500 MeV and the profiles
have been  averaged over all ecliptic longitudes. $(a)$ Profile derived with
no masking of Galactic diffuse emission or \gray{} point sources.  
$(b)$ Profile with the Galactic plane
($|b| < 10^\circ$ for $|l| > 90^\circ$ and $|b| < 20^\circ$ for $|l| <
90^\circ$) excluded.  $(c)$ Profile with the identified 3EG sources \citep{Hartman1999} 
and the Galactic plane excluded.  $(d)$ Profile with the identified 3EG sources plus
the further blazar identifications proposed by \citet{Sowards2003,Sowards2004}
excluded.  Overlaid on each profile is the best-fitting gaussian
(12.5$^\circ$ FWHM, centered on $\beta$ = 0) plus a constant, fit for the
region $|\beta| < 50^\circ$.  This approximates the distribution of albedo
\gray{} emission expected for the the KBO.
}
\label{fig:profiles}
\end{figure}

\section{Analysis of the EGRET data}\label{egret}

The EGRET instrument on the Compton Gamma-Ray Observatory (1991--2000)
surveyed the sky in the range $>$30 MeV and here we use the EGRET data
together with the information in the Third EGRET Source Catalog
\citep[3EG,][]{Hartman1999} to set limits on the signal from the SSSBs.
Challenges to detecting diffuse emission associated with the ecliptic
plane include the brightness of the Galactic diffuse emission 
\citep[e.g.,][]{Hunter1997}, the presence of bright point sources, 
the limited angular
resolution and photon statistics of the EGRET data, and potential
large-scale artifacts in the exposure maps owing to ageing of the spark chamber
gas.

We made maps of the EGRET data in ecliptic coordinates for Cycles 1--4 of
the mission, during which most of the EGRET exposure was obtained.  
The event
data, after standard cuts on zenith angle and inclination angle
(30$^\circ$), were binned on a photon-by-photon basis in ecliptic
coordinates.  
The exposure
maps for each EGRET viewing period were transformed into ecliptic
coordinates and added together and intensity maps were calculated from the
photon and exposure maps.

\placetable{Table1}

\begin{deluxetable*}{ccccc}
\tablecolumns{5}
\tablewidth{0pc}
\tabletypesize{\footnotesize}
\tablecaption{Diffuse intensity around the ecliptic (100--500 MeV) \label{Table1}}
\tablehead{
\colhead{Set of cuts} &
\multicolumn{2}{c}{$|\beta| < 15^\circ$}
\\ 
\colhead{in Figure~\ref{fig:profiles}} &
\colhead{Flux, cm$^{-2}$ s$^{-1}$} &
\colhead{Stat.\ error} & 
\colhead{Fitted flux, cm$^{-2}$ s$^{-1}$} &
\colhead{Stat.\ error} 
}
\startdata
$a$ & $1.006\times10^{-5}$ & $5.5\times10^{-7}$ & $9.16\times10^{-6}$ & $3.5\times10^{-7}$\\
$b$ & $7.95\times10^{-6}$  & $5.8\times10^{-7}$ & $5.95\times10^{-6}$ & $3.7\times10^{-7}$\\
$c$ & $3.59\times10^{-6}$  & $6.7\times10^{-7}$ & $3.53\times10^{-6}$ & $4.4\times10^{-7}$\\
$d$ & $1.1\times10^{-7}$   & $7.4\times10^{-7}$ & $1.52\times10^{-6}$ & $5.1\times10^{-7}$ 
\enddata
\end{deluxetable*}

In order to limit contributions from Galactic diffuse emission to any
possible enhancement of diffuse intensity at low ecliptic latitudes, the
region $|b| < 20^\circ$ for $|l| < 90^\circ$ and 
$|b| < 10^\circ$ for $|l| > 90^\circ$ was masked out in the analysis. 
We also
removed regions $12^\circ$ in diameter around the position of each identified 
source in the 3EG catalog (designation other than ``u'' in the 
3EG catalog).

Figure \ref{fig:profiles} presents the profile of \gray{} intensity in the 
100--500 MeV range over ecliptic longitude.  
This range was chosen as having the
brightest expected albedo emission in the energy range of EGRET.  
As described
in the caption, a sequence of profiles is shown for different combinations of
the masks described above.  
In the last profile (Figure~\ref{fig:profiles}$d$), the 3EG sources
subsequently identified by \citet{Sowards2003,Sowards2004} as likely to be
blazars were included with the sources identified in the 3EG catalog in
defining the mask.  
We did not mask out unidentified point sources because of
the possibility that some of them may have represented detections of the 
\gray{} albedo from the Trojan groups, which move collectively, or 
fluctuations in the SSSB \gray{} albedo at low ecliptic latitudes.

In order to estimate the possible ``excess'' diffuse flux from 
MBAs, Trojans, and KBOs 
we calculated the integrated fluxes for ecliptic latitudes
$|\beta|<15^\circ$ and all ecliptic longitudes (Table~\ref{Table1}, 
``Flux'' column).  
In order to increase the sensitivity, 
and to search for a diffuse 
signal that is centered
on the ecliptic the table also includes fluxes for the best-fitting Gaussian
centered on $\beta=0^\circ$ and having FWHM width 12.5$^\circ$ 
(``Fitted flux'' column),
the approximate extent of the Kuiper Belt \citep{Brown2001}.
The fits included a
constant term to account for Galactic and extragalactic diffuse emission;
profiles of the fits are included in Figure~\ref{fig:profiles}.  
The effective PSF for EGRET in
the 100--500 MeV range for the expected spectrum of the albedo emission is
approximately 4$^\circ$ FWHM, which would not appreciably broaden the apparent
distribution of \gray{} intensity.  
In any case, the assumption of a single
Gaussian profile is an approximation; the contribution from MBAs
should result in an additional somewhat narrower but fainter 
component to the diffuse emission around the ecliptic.

Gamma-ray emission associated with the Moon and Sun
also contributes
to the intensity of the sky at low ecliptic latitudes.  
The Moon is always
within about 5$^\circ$ of the ecliptic and the profiles shown in Figure 5
undoubtedly include lunar albedo \gray{} emission, at a level of $\sim5
\times 10^{-7}$ cm$^{-2}$ s$^{-1}$ \citep{MP2007b}.  
As described
by \citet{MPD2006} the solar radiation field is a fairly
bright and diffuse \gray{} source from inverse Compton scattering of
CR electrons.  
The solar inverse Compton emission is brightest in the
ecliptic plane but of course depends on solar elongation angle.  
The precise
contribution to the diffuse intensity at low ecliptic latitudes is difficult
to estimate. 
The Sun was in the field of view of EGRET for only a small
fraction of the observing time and the contribution to the total flux should
have been less than that of the Moon.

After the bright diffuse emission and identified point sources are masked from
the EGRET data, no strong excess of diffuse emission is apparent at low
ecliptic latitudes in Figure~\ref{fig:profiles}.  
The integrated fluxes are formally
significant for the case where the Galactic plane and all sources
identified in the 3EG catalog are masked out (Table~\ref{Table1}, 
Figure~\ref{fig:profiles}$c$), but the systematic
uncertainties are comparable to the measurement.  
This is suggested by the
effect on the integrated flux from masking out several more 
sources that
\citet{Sowards2003,Sowards2004} identified as 
blazars (Figure~\ref{fig:profiles}$d$).  
The
overall average exposure does not change appreciably as a result of the
additional masking but the fit flux decreased by more than 
50\% (Table~\ref{Table1}).

Based on our analysis of the EGRET data we infer that the diffuse emission
from MBAs, Trojans, and KBOs
has an integrated flux of less than $\sim$$6
\times 10^{-6}$ cm$^{-2}$ s$^{-1}$ (100--500 MeV), as derived from the set of
cuts $b$, which corresponds to $\sim$12 Lunar albedo flux units. 


\section{Discussion and Conclusion}

The albedo \gray{} flux from MBAs
can be calculated using eqs.~(\ref{eq7})--(\ref{eq11})
and Figures~\ref{fig:moon1}, \ref{fig:moon}
where we assume that their surface material is regolith.
We use the following parameters:
$\rho=2$ g cm$^{-3}$ for the MBA average density, 
$r_1=4.565\times10^7$ cm for the radius of
Ceres,
$r_0=100$ cm for the smallest radius of an asteroid that is still 
an opaque target for incident CR particles.
The central grammage in this case ($r_0=100$ cm)
is $\sim$400 g cm$^{-2}$. Since the composition of the MBAs 
(and other SSSB populations) is 
mainly oxygen, this corresponds to $\sim$5 interaction lengths which is 
sufficient 
for the hadronic cascade to fully develop at the CR energies we consider.
With these parameters, the total MBA
albedo flux is 
${\cal X}=F_{\rm tot}/F_{\moon}=0.05$, 0.67, 12
for extrapolation to small sizes with indices $n=2.5$, 3.0, 3.5 
(see Figure~\ref{fig1a}), correspondingly.

Similarly, for the Jovian Trojan asteroids we can estimate the \gray{} flux 
assuming the same size distribution as for MBAs,
but with $\ell\sim5.2$ AU, and $r_0=200$ cm (which gives the same
central grammage for $\rho=1$ g cm$^{-3}$). We obtain
${\cal X} = F_{\rm tot}/F_{\moon} = 0.01$, 0.07, 0.8 
(averaged over their orbit) for a similar 
extrapolation to small sizes with indices $n = 2.5$, 3.0, 3.5.
For the closest (4.2 AU) and the farthest (6.2 AU) distances to Earth,
the fluxes will be 
0.01, 0.1, 1.2 and 0.006, 0.05, 0.5, correspondingly.

The KBO size distribution is known very approximately.
The second largest object of the Kuiper Belt after 136199 Eris is 
Pluto $r_1=1.195\times10^8$ cm,
while the majority of the KBOs are icy rocks and comets
with $\rho=0.5$ g cm$^{-3}$. 
To keep the same central grammage of the smallest body 
$\sim$400 g cm$^{-2}$ we have to use a larger minimum radius, $r_0=400$ cm, than
for the MBAs.
The incident spectrum of CR particles at $>$30 AU approaches the
LIS spectrum which results in a factor of $\sim$2 increase below $\sim$1 GeV
of the albedo flux compared to the same body at 1 AU 
(Figure~\ref{fig:moon1}).
For these parameters, the total Kuiper Belt albedo flux is 
${\cal X}^{\rm K}=F^{\rm K}_{\rm tot}/F_{\moon}=0.2$, 34, 1168
for $n=3.0$, 3.5, 3.9, correspondingly. 
The removal of Eris and Pluto
($\sim$$0.005M_\oplus$ combined) from the Kuiper Belt and using Charon
instead, $r_1=6\times10^7$ cm, results in the flux increase:
${\cal X}^{\rm K}=F^{\rm K}_{\rm tot}/F_{\moon}=0.35$, 46, 1222
for the same values of $n$.
However, this change is simply the result of anchoring the power-law
size distribution to a large body.

Our estimates show that the albedo of MBAs and KBOs 
could
account for the EGRET upper limit of the flux from the ecliptic.
For the adopted size distributions of SSSBs 
($n=3.0$ for MBAs and $n=3.5$ for KBOs), the KBO albedo is essentially 
brighter. 
However, if the MBA size distribution is somewhat steeper
than our adopted index of $n = 3.0$, e.g., as for the distribution 
proposed by \citet{Cheng2004}, it can
account for the total albedo flux from the ecliptic.
The SSSB \gray{} albedo, especially of the collectively moving Trojan groups, 
might be responsible for some fraction of the EGRET unidentified point 
sources at low ecliptic latitudes.

A possible way to distinguish the albedo emission of MBAs and KBOs
is to study the emission as a function of solar elongation angle. 
In the antisolar direction, $\theta\approx180^\circ$, the direction in which 
the Main Belt is closest to the Earth ($\sim$1.7 AU), the flux is predicted 
to be as much as $\sim$5 times that in the solar direction, 
$\theta\approx0^\circ$.  
On the other hand, the brightness of the Kuiper Belt is only weakly 
dependent on the elongation angle because it is much further away.
The positions of the Trojans on the sky are known, being 
determined relative to their respective planet (Jupiter, Neptune), 
making them easier to detect.

The detection of the CR-induced \gray{} albedo of MBAs, Trojans, and 
the KBOs by \gray{} instruments is possible. 
At higher energies $\ga$1 GeV where the \gray{} albedo flux is steady
and does not depend on the solar modulation, it can serve as
a normalization point to the cumulative brightness 
of all MBAs plus KBOs.
At lower energies $\la$1 GeV, 
the component of the albedo which is independent of elongation,
the KBO albedo,
will tell us directly about the LIS spectrum of CRs. 
Therefore, the observations of the albedo flux can provide 
us with valuable information about the size distributions of SSSBs in both
regions, while the shape of the albedo spectrum can tell us about
the LIS spectra of CR protons and helium at high energies.
In turn, a detection of the MBA and KBO albedo
at MeV-GeV energies will enable us to normalize properly the cumulative
albedo spectrum and make a prediction for the intensity of the 511 keV line.

A conservative estimate of the 511 keV line flux from SSSBs can be made using
the upper limit derived in Section~\ref{egret}.
The total flux of 511 keV
photons from the Moon is $F_{\moon}^{511}
\approx10^{-3}\Delta E\approx 2.4\times10^{-5}$ photons cm$^{-2}$
s$^{-1}$ (Figure~\ref{fig:moon1}, $\Phi=0$ MV), 
where $\Delta E=0.024$ MeV is the size of the bin containing $E=0.511$ MeV. 
The total flux from the SSSBs
${\cal X}^{511}=F_{\rm tot}^{511}/F_{\moon}^{511}$ 
can be calculated from eqs.~(\ref{eq7})--(\ref{eq11}).
The SSSB albedo contribution to the 511 keV line flux within 
the Galactic bulge is 
$\sim0.72 F_{\moon}^{511}$,
where we assumed that the FWHM
of the bulge is $\sim$$10^\circ$ \citep{Knodlseder2005}, and $20^\circ/360^\circ\approx 0.06$,
and we used the upper limit 
derived in Section~\ref{egret}.
It
gives $\sim$$2\times10^{-5}$
photons cm$^{-2}$ s$^{-1}$, which is about 2\% of the total bulge emission as 
observed by the INTEGRAL $(1.05\pm0.06)\times10^{-3}$ photons cm$^{-2}$ s$^{-1}$ 
\citep[][]{Knodlseder2005}.
Since most of the INTEGRAL observing time was spent on observations of the Galactic bulge
and a comparatively small fraction went into observing regions 
above and below the Galactic plane,
it is not surprising that a diffuse band near the ecliptic (the SSSB albedo)
has not been found so far. It is interesting that 
the OSSE map of the 511 keV line has a controversial feature, the so-called
``annihilation fountain,'' above the Galactic bulge
\citep{Purcell1997} which, in fact, may be the asteroid albedo foreground instead.
Note, that the
\gray{} spectrometer on the NEAR-Shoemaker spacecraft made observations
of the 511 keV line from 433 Eros \citep{Evans2001} on the surface of 
Eros itself, 
however, it is hard to judge the absolute intensity of the line from the 
published data.

Our estimates of the fluxes assume that the mass and radius distributions
are valid for the whole range of masses, which is not necessarily true. 
The number of small bodies may be larger or smaller than the extrapolation
from the distribution of more massive bodies. 
We have also assumed spherical bodies. 
However, the smallest bodies are distinctly non-spherical which would make 
them somewhat brighter than we have estimated. 
Thus, our calculations underestimate the SSSB albedo emission.

The bodies that are smaller than the 
cutoff radius 
($r_0=100$ cm for MBAs, 200 cm for Trojans, 400 cm for KBOs)
will also contribute to the albedo flux.
Because of their smaller size, only the initial stage of the CR 
cascade will develop, producing a harder albedo spectrum while
its intensity will be reduced due to the partial
conversion of energy of the primary CR particles into albedo \gray{s}.

We emphasise that the detection of the \gray{} albedo from MBAs, KBOs, and 
other SSSB families directly probes the size distribution of these bodies 
below the detection limit of other methods, over considerably larger regions
of the sky.
The detectability of the \gray{} emission by these objects has implications
for studies of 
the evolution of the
solar and exo-solar planetary systems \citep{Brown2004}, 
studies of CRs,
and diffuse \gray{s}.
The GLAST Large Area Telescope (LAT)\footnote{See the GLAST LAT performance Web-page: {\tt http://www-glast.slac.stanford.edu}}, 
to be launched by NASA in May 2008, will in just one year 
have an essentially uniform exposure over the entire sky 
a factor of 40 or more deeper than EGRET
and will be free from sensitivity variations owing to ageing of consumables.
This capability will permit detection of albedo \gray{} fluxes for SSSBs at
even the Lunar flux level.

\acknowledgments
I.\ V.\ M.\ wishes to dedicate this paper to the memory of his mother. 
We thank Clark Chapman and Bill Merline for careful 
reading of the manuscript and insightful 
remarks, and Joe Burns, Alan Harris, and Ed Tedesco 
for sharing their thoughts.
This investigation was inspired by a question
from NASA Associate Administrator for the Science Mission Directorate, Dr.\
S.\ Alan Stern.
I.\ V.\ M.\ and J.\ F.\ O.\ acknowledge support from NASA
Astronomy and Physics Research and Analysis Program (APRA) grant.
T.\ A.\ P.\ acknowledges partial support from the US Department of Energy.
P.\ F.\ M.\ acknowledges support from NASA contract NAS5-00147 for GLAST.
This work was carried out while J.\ F.\ O.\ was a visiting scientist at Stanford 
University; he wishes to acknowledge the kind hospitality.
This research has made use of data obtained through the High Energy Astrophysics 
Science Archive Research Center, 
provided by the NASA/Goddard Space Flight Center.

\end{document}